\documentstyle[prl,aps,twocolumn,epsf]{revtex}
\begin{document}
\draft
\title{PERTURBATION THEORY FOR CLASSICAL SOLIDS}

\author{C. Rasc\'{o}n$^{1,2}$, L. Mederos$^2$, and G. Navascu\'{e}s$^1$}
\address{$^1$Departamento de F\'{\i}sica Te\'{o}rica de la Materia
Condensada, Universidad Aut\'{o}noma, Cantoblanco, Madrid E-28049,Spain}
\address{$^2$Instituto de Ciencia de Materiales (Consejo Superior de
Investigaciones Cient\'{\i}ficas), Cantoblanco, Madrid E-28049, Spain}
\date{\today}
\maketitle

{\bf Phys.Rev.Lett.77, p.2249 (1996)}

\begin{abstract}
The first well founded perturbation theory for classical solid
systems is presented. Theoretical approaches to thermodynamic
and structural properties of the hard-sphere solid provide us 
with the reference system. 
The traditional difficulties of all previous approaches are overcome.
The perturbation 
is a first order term in an expansion of the Helmholtz free energy
functional instead of an additive {\it ad hoc} term and
the proper solid reference structure is used instead of some kind of
mapping into the fluid structure at some effective density. 
As the theory reduces to the usual liquid perturbation theory
in the uniform limit, it can describe consistently the complete phase diagram.
Excellent results are obtained when applied to different model systems.
\end{abstract}

In the last two decades, we have witnessed a 
continuous progress in the theory of
non uniform classical fluids \cite{HM,Hend}. Inside this context, where
 classical
solids are regarded as an extremely non uniform system, a considerable
effort has been devoted to develop a theory for solids. From this
point of view, the theory would provide not only a description of 
solids but a unified view of classical systems \cite{Hend}. 
However, due to the inherent difficulties in the
non uniform systems, this progress is still not comparable to that
reached in uniform liquids.

An important part of the enormous progress of the theory of simple uniform 
liquids has been due to the
development of perturbation theories. Within this scheme,
the liquid properties
are described by those of a reference system (usually built with the
repulsive part of the interaction potential) modified by a perturbation
(usually due to the attractive part of the interaction potential).  
The usefulness of these theories is based
on the knowledge of the thermodynamic and structural properties of the reference
system. The foremost, if not the unique, reference system 
is the hard-sphere fluid, which describes many of the essential
features of realistic interacting repulsive potentials. For this reason,
much effort has been devoted to understanding its properties.
Nowadays, the virial expansion, the analytically
solvable Perkus-Yevick approximation and semi-empirical
approximations supply quite accurate results for thermodynamics
and structure up to the crystallization density \cite{HM}.
The seminal article by
Andersen {\it et al.} \cite{WCA} establishes the relation between the
properties of a hard-sphere fluid and those of realistic repulsive
interactions, completing the scheme of the perturbation theory for simple
liquids. 

The same strategy has been pursued in the solid phase of classical systems.
The aims of the theory are essentially two: the formal development
of the theory itself and the determination of accurate thermodynamic
and structural properties of the reference system. 
To this end, the density functional formalism has been
successfully elaborated to describe the {\it thermodynamic} properties of the
hard-sphere solid \cite{Evans92}. At present, there are different
functional approaches for the Helmholtz free energy which describe the solid
phase and reduce to that of the fluid in the homogeneous limit
\cite{Hend}. Recently, it has been showed that the equation of state obtained
from these functionals reproduces the simulation results quite well even up to
almost close packing \cite{EoS}. For all that, 
the thermodynamics of the hard-sphere
system can be accurately described from low densities to almost close
packing, including the crystallization phase transition.

However, up to very recently, there was neither an accessible 
theoretical approach to the structural properties of the hard-sphere solid 
nor a proper perturbation theory \cite{GTILDE}. All the perturbation approaches 
proposed by different authors, two of us included,
are rather primitive \cite{teoper,KB,PWDA}.
Basically, they follow the same scheme: the
Helmholtz free energy of the solid is written as a sum of two terms, namely the
reference and perturbation free energies. The former is 
assumed to be any of the available
density functionals for the free energy of the hard-sphere solid 
while the
perturbation is an {\it ad hoc} term built
without any connection with the hard-sphere potential which is being assumed
as the reference potential. Furthermore, the structure of the reference 
hard-sphere solid is substituted in the perturbative term
by some kind of mapping into the structure of the hard-sphere 
{\it fluid} at some effective density.
Typically, the Andersen {\it et al.} (WCA) criterium is used to divide
the interacting potential, whereas the Barker-Henderson criterium
\cite{HM,BH} is used to fix the hard-sphere diameter of the reference
solid. To worsen things, the Barker-Henderson criterium does not
distinguish uniform from non uniform systems. 
Any intent to do a consistent approach gives disastrous results. 

In spite of the crudeness and the fragility of these approaches,
all pieces assembled achieved 
to describe the phase diagram of the Lennard-Jones system 
reasonably well \cite{KB}. This
success is not completely understood though some clues 
have been suggested \cite{PWDA}.
It seems that for large-ranged attractive interactions some kind 
of numerical error cancelation should occur. When
they are applied to systems with slightly more sophisticated potentials,
they merely give a qualitative description of the thermodynamic properties
or even fail completely. This is what happens
when short-ranged attractive
interactions are present. In any case, these theories are far from
being satisfactory.

The first step in order
to develop a proper theory of perturbations for solids has been
carried out recently by determining theoretically 
the average $\tilde g(r)$ of the pair distribution function
$\rho^{(2)}({\bf r}_1,{\bf r}_2)$ \cite{GTILDE}. This average is defined by

\begin{equation}
\tilde g(r_{12}) =  {{1} \over {4 \pi V \rho^2}}
\int d \Omega \int d{\bf r}_1 \rho^{(2)}({\bf r}_1,{\bf r}_2),
\label{defg}
\end{equation}
where $V$ is the volume, $\rho$ is the mean density, and $d \Omega$ the
differential solid angle aperture around ${\bf r}_{12}$. 
Excellent results are obtained for the
hard-sphere solid up to almost close packing.

In this Letter, we develop a perturbation theory where,
to first order, all the structural
information needed is precisely the function $\tilde g(r)$.
As it is usually done in perturbation theories for uniform liquids, we divide
the interacting potential into the reference and the perturbative parts:
$\phi(r)=\phi_r (r)+\phi_p (r)$. Then, we start from the general expression
for the Helmholtz free energy of a non uniform system as a functional of the
density $\rho ({\bf r})$ which is written exactly as \cite{Evans92,RW}

\begin{equation}
F[\rho({\bf r})]=F_{r}[\rho({\bf r})]+F_{p}[\rho({\bf r})],
\label{F}
\end{equation}
being $F_{r}[\rho({\bf r})]$ the Helmholtz free energy of the reference system
at density $\rho ({\bf r})$ and

\begin{equation}
F_{p}[\rho({\bf r})] = {{1} \over {2}}
\int_{0}^{1} d\alpha
\int d {\bf r}_1 d{\bf r}_2
\rho^{(2)}({\bf r}_1,{\bf r}_2;\alpha) \phi_p (r_{12}),
\label{Fp}
\end{equation}
where $\alpha$ is the coupling parameter for the interaction potential
$\phi(r;\alpha)=\phi_{r}(r) + \alpha \phi_{p}(r)$ and
$\rho^{(2)}({\bf r}_1,{\bf r}_2;\alpha)$ is the pair distribution
function when the potential
is $\phi(r_{12};\alpha)$ but the density is $\rho({\bf r})$. To lowest order,
Eq.(\ref{Fp}) yields

\begin{equation}
F_{p}[\rho({\bf r})] = {{1} \over {2}}
\int d {\bf r}_1 d{\bf r}_2 
\rho^{(2)}_{r}({\bf r}_1,{\bf r}_2) \phi_p (r_{12}),
\label{Fp1}
\end{equation}
which after an appropriate integration becomes exactly

\begin{equation}
F_{p}[\rho({\bf r})] = {2 \pi \rho N} 
\int dr {r}^2  \tilde g_{r}(r) {\phi}_{p}(r),
\label{Fp2}
\end{equation}
where $N$ is the number of particles and $\tilde g(r)$ is precisely the
average of the pair distribution function given by Eq.(\ref{defg}).
For any realistic system, the reference interacting potential is 
chosen to describe
the rapidly varying short-ranged repulsive part of the interacting potential.
However, the thermodynamic and structural properties of these systems are
unknown. Therefore, a treatment to relate these properties to those of
a hard-sphere solid is evidently necessary. The procedure is analogous to that
of the theory of liquids. Let $e({\bf r}_1,{\bf r}_2)
= exp(-\beta \phi |{\bf r}_1-{\bf r}_2|)$ be the Boltzmann factor. 
It depends only on
$r_{12}$, but for clearness in some of the expressions below it is
convenient to keep the formal dependence on the two locations
${\bf r}_1$ and ${\bf r}_2$. As in the liquid theory, let
$\Delta e({\bf r}_1,{\bf r}_2)$ be the blip function, {\it i.e.},
 the difference of the Boltzmann factors of the reference potential and
the hard-sphere potential of diameter $d_{HS}$, $e_r(|{\bf r}_1-{\bf
r}_2|)-e_{HS}(|{\bf r}_1-{\bf r}_2|)$. If this diameter is
comparable to
the range of the reference potential, the blip function
is different from zero in a small range of the order of $\xi
d_{HS}$ with $\xi<1$. We can then expand the Helmholtz free energy functional
of the reference system in powers of the blip function

\begin{eqnarray}
F_{r}[\rho({\bf r})] = F_{HS}[\rho({\bf r})] +
\nonumber \\
{{1} \over {2}}\int d{\bf r}_1 \int d{\bf r}_2 
{{\delta F_{HS}[\rho ({\bf r})]}\over{\delta e_{HS}({\bf r}_1,{\bf r}_2)}}
\Delta e({\bf r}_1,{\bf r}_2)  
\nonumber \\
+ \mbox{higher-order terms}.
\label{Fexp}
\end{eqnarray}
The functional derivative of the Helmholtz free energy 
with respect to the Boltzmann factor is easily obtained

 \begin{equation}
- \beta {{\delta F[\rho ({\bf r})]} \over 
{\delta e({\bf r}_1,{\bf r}_2)}}
= {{1} \over {2}} \rho({\bf r}_1) 
\rho({\bf r}_2) y({\bf r}_1,{\bf r}_2),
\label{y}
\end{equation}
where the $y({\bf r}_1,{\bf r}_2)$ function
is the generalization of the $y(r_{12})$ function in uniform systems.  This
last one is defined by $e^{-\beta\phi(r_{12})} g(r_{12})$, whereas the former
is defined by $e({\bf r}_1,{\bf r}_2) { {\rho^{(2)}({\bf r}_1,{\bf r}_2)}
\over {\rho({\bf r}_1) \rho({\bf r}_2)} }$ .
Introducing Eq.(\ref{y}) into Eq.(\ref{Fexp}) and after some simple
algebraic steps, it is found that

\begin{eqnarray}
F_{r}[\rho({\bf r})] = F_{HS}[\rho({\bf r})] +
\nonumber \\
2 \pi {\rho} N \int dr r^2 {\tilde y}_{HS}(r)
\Delta e(r)
+ \mbox{higher-order terms},
\label{Fexp2}
\end{eqnarray}
where $\tilde y_{HS}(r)$ is the average of the $y_{HS}({\bf r}_1,{\bf
r}_2)$ function which can also be expressed as $\tilde y_{HS}(r)=exp(\beta
\phi_{HS}(r)) \tilde g_{HS}(r)$. Until here, the hard-sphere diameter has
remained arbitrary. Now we specify it by imposing that the first order term in
the functional expansion Eq.(\ref{Fexp2}) be zero. This yields

\begin{equation}
\int_{d_{HS}}^{\infty} dr r^{2} {\tilde y}_{HS}(r) =
\int_{0}^{\infty} dr r^{2} {\tilde y}_{HS}(r) exp(\beta \phi_{r}(r)).
\label{WCA}
\end{equation}
Notice that it is analogous to the well known WCA criterium \cite{WCA} in
liquid theory except that the $\tilde y(r)$ function instead 
of the $y(r)$ function
is used. From Eq.(\ref{Fexp}) and Eq.(\ref{y}) and after 
averaging, we find

\begin{equation}
{\tilde y}_r(r) = {\tilde y}_{HS}(r) + \mbox{higher-order terms},
\label{y0}
\end{equation}
which can be used to lowest order to evaluate the perturbation term
Eq.(\ref{Fp2}).
It is interesting and easy to prove that, if Eq.(\ref{WCA}) holds,
the convergence properties of the expansions 
(\ref{Fexp2}) and (\ref{y0}) are similar to those found
in the uniform limit \cite{WCA2}, {\it i.e.}

\begin{eqnarray}
F_{p}[\rho({\bf r})] = F_{HS}[\rho({\bf r})] +
\nonumber \\
2 \pi \rho N \int dr r^2 {\tilde y}_{HS}(r)
\Delta e(r)
+ O(\xi ^4),
\label{Fexp3}
\end{eqnarray}
and 

\begin{equation}
{\tilde y}_r(r) = {\tilde y}_{HS}(r)+O(\xi^2).
\label{y02}
\end{equation}

With the above expressions the theoretical formalism is completed. It
is straightforward to see that in the uniform limit the 
theory reduces to the WCA perturbation theory of simple liquids. 
Observe also that all the problems of
previous approaches are automatically overcome: the formalism provides
a proper expansion of the free energy
with precise thermodynamic and structural properties of
a reference solid, namely any of the functionals for the Helmholtz free energy
and the correlation function $\tilde g(r)$, and the criterium to
determine the appropriate reference system enhances the convergence of
the expansion.

We have applied the theory to the Lennard-Jones (LJ) system and to an 
extremely short-ranged square well potential. In Table I we show the 
LJ liquid ($\rho_l$) and solid ($\rho_s$) densities at
coexistence at several temperatures (all in LJ parameters units) obtained
from the present theory, using Tarazona functional \cite{T} for the
reference hard-sphere solid, compared to simulation data of Hansen and
Verlet \cite{HV}. The
theoretical predictions are quite good and the small deviations are
quantitatively similar to those obtained by previous {\it ad hoc}
approaches. Nevertheless, contrary to what it is predicted by the later, the
present theory gives the correct slopes of the coexistence densities
as functions of the temperature. This improvement is undoubtfully due to 
the adequate dependence of the hard-sphere diameter with the density.
Table I also shows the solid Lindemann parameter at coexistence.
It is significantly smaller than the simulation result.
This is due to the functional approach we have used for the reference
hard-sphere solid, as it is well known. If any other
functional were used, the Lindemann parameter would improve but
no significant changes would be detected for the energies and, therefore,
for the coexistence densities shown in Table I.
However, the important point is that,
for the first time, a classical system can be studied with a
unique, proper and consistent theory. 

Much more impressive are the results obtained
by the theory when it is applied to a system with 
an extremely short-ranged attractive
interaction. 
Simulation results have recently proved \cite{Frenkel} that this kind of
systems present 
quite interesting isostructural solid-solid transitions. Up to now,
as one could expect, no theory had been able to give a reasonable
quantitative account of this behavior \cite{SQ,LicTej}.
Therefore, these systems offer an interesting and stringent
test for the present theory. Figure 1 shows the phase diagram of the
square well potential characterized by a short width
 $\delta/d_{HS}=0.02$.
The dashed lines of this figure, which corresponds to
simulation results \cite{Frenkel}, show the coexistence densities of 
the liquid-solid transition and, inside of the solid region,
the coexistence curve 
of the solid-solid transition. Observe that, due to
the weakness of the attractive potential, there is no fluid condensation.
However, the solid-solid transition is a kind of solid condensation
which arises from a different mechanism than the usual fluid
condensation. It is related with the commensurability of the
lattice parameter of the solid structure with the well width of the potential
\cite{Frenkel,SQ}. The dotted lines correspond to the
perturbation weighted density approximation (PWDA) \cite{PWDA,SQ}
which was, to our knowledge, the best approach so far able to describe
the complete phase diagram of this system.
The continuous lines
correspond to the present theory using Tarazona functional \cite{T} for the
reference hard-sphere solid. The results using any other functional would
be indistinguishable in the scale of this figure. The dramatic improvement
is clear and the global agreement of the theoretical predictions
with the simulation results is quite good. Equivalent quantitative
agreement is obtained for the phase diagrams corresponding to square
wells of different widths.

In summary, we have developed a perturbation theory 
with remarkable characteristics. It is the
first well founded perturbation theory for classical solids. 
It reduces to the well
known WCA perturbation liquid theory in the appropriate uniform limit. 
Its versatility allows its use with 
any of the functional approaches available for the Helmholtz free
energy of hard spheres. It gives quite good 
results even for systems where other {\it ad hoc} approaches have failed.
Finally, but not less important,
the computational effort is drastically reduced. The cumbersome mapping of
the structure of the solid into the structure of the liquid at certain
effective density, which usually involves tedious autoconsistent and
recursive processes, is replaced by a simple integral.

This work has been supported by the Direcci\'{o}n General de Investigaci\'{o}n
Cient\'{i}fica
y T\'{e}cnica of Spain under grant number PB94-0005-C02.

\begin{figure}
\caption{
Phase diagram of the square well system for the well width 
$\delta/d_{HS}=0.02$. The temperature is given in well depth units.
The solid line is the prediction of the present theory. Dashed line
corresponds to the simulations by Bolhuis and Frenkel. Dotted line is the
prediction of the PWDA.}
\end{figure}

\begin{table}
\caption{
Lennard-Jones liquid ($\rho_l$) and solid ($\rho_s$) densities
at coexistence
at several temperatures (all in Lennard-Jones parameters units) as 
predicted by the simulations of Hansen and Verlet and the present theory.
The Lindemann parameter L of the solid phase at coexistence is also shown.}
\begin{tabular}{ccc}
\\
        & Simulation              &  Theory                            \\
$k_BT$&$\rho_l$~~~~~~~$\rho_s$~~~~~~~L&$\rho_l$~~~~~~~$\rho_s$~~~~~~~L \\ 
\\
\hline
\\
   0.75 & 0.875~~~~0.973~~~~0.145   &  0.884~~~~0.970~~~~0.087      \\  \\
   1.15 & 0.936~~~~1.024~~~~0.139   &  0.974~~~~1.049~~~~0.082      \\  \\
   1.35 & 0.964~~~~1.053~~~~0.137   &  0.996~~~~1.077~~~~0.083      \\  \\
   2.74 & 1.113~~~~1.179~~~~0.140   &  1.116~~~~1.199~~~~0.090      \\  \\
\end{tabular}
\end{table}


\begin{references}
\bibitem{HM} J.P. Hansen and I.R. McDonald, {\it Theory of Simple
Liquids} (Academic, New York, 1986).
\bibitem{Hend} {\it Fundamentals of Inhomogeneous Fluids}, edited by D.
Henderson (Dekker, New York, 1992)
\bibitem{WCA} H. C. Andersen, J. D. Weeks, and D. Chandler, Phys. Rev. A
{\bf 4}, 1597 (1971).
\bibitem{Evans92} For a recent review see, for example, R. Evans, in {\it
Fundamentals of Inhomogeneous Fluids}, edited by D. Henderson (Dekker, New
York, 1992), Chapter 3, and references therein.
\bibitem{EoS}A. R. Denton, N. W. Ashcroft, and W. A. Curtin, Phys. Rev. E
{\bf 51}, 65 (1995).
C. F. Tejero, M. S. Ripoll, and A. P\'{e}rez, Phys. Rev. E
{\bf 52}, 3632 (1995).
C. Rasc\'{o}n, G. Navascu\'{e}s and L. Mederos, Phys.
Rev E {\bf 53}, 5698 (1996).
\bibitem{GTILDE} C. Rasc\'{o}n, L. Mederos and G. Navascu\'{e}s, Phys.
Rev E {\bf 54}, 1261 (1996).
\bibitem{teoper} 
W. A. Curtin, N. W. Ashcroft, Phys. Rev. Lett. {\bf 56}, 2775 (1986), 
Z. Tang, L. E. Scriven and H. T. Davis, J. Chem. Phys. {\bf 95}, 2659 (1991), 
S. Sokolowski and J. Fischer, J. Chem. Phys. {\bf 96}, 5441 (1992), 
L. Mederos, G. Navascu\'{e}s, P. Tarazona and
E. Chac\'{o}n, Phys. Rev. E {\bf 47}, 4284 (1993).
\bibitem{KB} A. Kyrlidis and R. A. Brown, Phys. Rev. E {\bf 47}, 427 (1993).
\bibitem{PWDA} L. Mederos, G. Navascu\'{e}s, and P. Tarazona, Phys. Rev.
E {\bf 49}, 2161 (1994).
\bibitem{BH} J. A. Barker and D. Henderson, J. Chem. Phys. {\bf 47}, 4714
(1967).
\bibitem{RW} J. Rowlison and B. Widom, 
{\it Molecular Theory of Capilarity} (Clarendon, Oxford, 1984).
\bibitem{WCA2} H. C. Andersen, J. D. Weeks, and D. Chandler, J. Chem.
Phys. {\bf 54}, 5237 (1971).
\bibitem{T} P. Tarazona, Mol. Phys. {\bf 52}, 81 (1984) and 
Phys. Rev. A {\bf 31}, 2672 (1985).
\bibitem{HV} J. P. Hansen and L. Verlet, Phys. Rev. {\bf 184}, 151 (1969).
\bibitem{Frenkel} P. Bolhuis and D. Frenkel, Phys. Rev. Lett. {\bf 72}, 2211
(1994).
\bibitem{SQ} C. Rasc\'{o}n, G. Navascu\'{e}s, and L. Mederos, Phys. Rev. B
{\bf 51}, 14899 (1995). 
\bibitem{LicTej} C. F. Tejero, A. Daanoun, H. N. W. Lekkerkerker, and M.
Baus, Phys. Rev. Lett. {\bf 73}, 752 (1994), C. N. Likos, Zs. T. N\'emeth,
and H. L\"{o}wen, J. Phys.: Condens. Matter, {\bf 6}, 10965 (199494)
\end{references}
\end{document}